\documentclass{article}
\usepackage{spconf,amsmath,graphicx}
\usepackage[table,xcdraw]{xcolor}
\usepackage{algorithmic}
\usepackage{algorithm}
\usepackage{array}
\usepackage{textcomp}
\usepackage{stfloats}
\usepackage{verbatim}
\usepackage{graphicx}
\usepackage{cite}
\usepackage{pifont}
\newcommand{\xmark}{\ding{55}}
\usepackage{adjustbox}
\newcommand{\etal}{\emph{et al.}~}
\usepackage{hhline}
\usepackage{adjustbox}
\usepackage{float}
\usepackage{booktabs}
\usepackage{multirow}
\usepackage{tabu}
\usepackage{lscape}
\usepackage[normalem]{ulem}
\usepackage{arydshln}

\usepackage{lettrine}
\usepackage{verbatim}
\usepackage[breaklinks]{hyperref}
\usepackage{url}

\usepackage{lscape}
\usepackage{comment, soul}
\usepackage{bm}
\usepackage{subcaption}
\usepackage{diagbox}
\usepackage{slashbox}
\usepackage{makecell}

\usepackage{caption}
\usepackage{cite}
\usepackage{paralist}
\usepackage{amsmath,amssymb,amsfonts}
\usepackage{algorithmic}
\usepackage{graphicx}
\usepackage{subcaption}
\usepackage{threeparttable}
\usepackage{textcomp}
\usepackage{xcolor}
\usepackage{tikz}
\def\BibTeX{{\rm B\kern-.05em{\sc i\kern-.025em b}\kern-.08em
    T\kern-.1667em\lower.7ex\hbox{E}\kern-.125emX}}
\usepackage{acronym}
\newacro{has}[HAS]{HTTP adaptive streaming}
\newacro{uav}[UAV]{unmanned aerial vehicle}
\newacro{vr}[VR]{virtual reality}
\newacro{xr}[XR]{extended reality}
\newacro{vista}[ODVista]{omnidirectional video streaming dataset}
\newacro{ra}[RA]{random access}
\newacro{sr}[SR]{super-resolution}
\newacro{hevc}[HEVC]{high-efficiency video coding}
\newacro{odv}[ODV]{omnidirectional video}
\newacro{cc}[CC]{creative commons attribution}
\newacro{erp}[ERP]{equirectangular projection}
\newacro{si}[SI]{spatial information}
\newacro{ti}[TI]{temporal information}
\newacro{cf}[CF]{colorfulness}
\newacro{vca}[VCA]{video complexity analyzer}
\newacro{br}[BR]{brightness}
\newacro{cnn}[CNN]{convolution neural network}
\newacro{rnn}[RNN]{recurrent neural network}
\newacro{gan}[GAN]{generative adversarial network}
\newacro{vit}[ViT]{vision transformer}
\newacro{lr}[LR]{low resolution}
\newacro{hr}[HR]{high resolution}
\newacro{ra}[RA]{random access}
\newacro{gpu}[GPU]{graphics processing units}
\newacro{ws-psnr}[WS-PSNR]{weighted spherical peak signal-to-noise ratio}
\newacro{ws-ssim}[WS-SSIM]{weighted spherical structural similarity index measure}
\newacro{gpu}[GPU]{graphics processing unit}
\newacro{ai}[AI]{artificial intelligence}
\newacro{fps}[fps]{frames per second}
\newacro{ml}[ML]{machine learning}
\newacro{vod}[VoD]{video-on-demand}
\newacro{fsrcnn}[FSRCNN]{fast super-resolution convolutional neural network}
\newacro{vvc}[VVC]{versatile video coding}

\usepackage{hyperref}

\hypersetup{
    colorlinks=true,
    linkcolor=blue,
    filecolor=magenta,      
    urlcolor=cyan,
} 
\title{ODVista: An Omnidirectional Video Dataset for super-resolution and Quality Enhancement Tasks}

\name{Ahmed Telili$^{1}$, Ibrahim Farhat$^{1}$, Wassim Hamidouche$^{1}$, Hadi Amirpour$^{2}$}
\address{$^{1}$ Technology Innovation Institute P.O.Box: 9639, Masdar City, Abu Dhabi, UAE
      \\
      $^{2}$Christian Doppler Laboratory ATHENA, Alpen-Adria-Universit{\"a}t, Klagenfurt, Austria
      \vspace{-0.15in}}
\begin{document}
\ninept
\maketitle

\begin{abstract}
Omnidirectional or 360-degree video is being increasingly deployed, largely due to the latest advancements in immersive \ac{vr} and \ac{xr} technology. However, the adoption of these videos in streaming encounters challenges related to bandwidth and latency, particularly in mobility conditions such as with \acp{uav}. Adaptive resolution and compression aim to preserve quality while maintaining low latency under these constraints, yet downscaling and encoding can still degrade quality and introduce artifacts. \Ac{ml}-based \ac{sr} and quality enhancement techniques offer a promising solution by enhancing detail recovery and reducing compression artifacts. However, current publicly available 360-degree video \ac{sr} datasets lack compression artifacts, which limit research in this field. To bridge this gap, this paper introduces \ac{vista}, which comprises 200 high-resolution and high-quality videos downscaled and encoded at four bitrate ranges using the \ac{hevc}/H.265 standard. Evaluations show that the dataset not only features a wide variety of scenes but also spans different levels of content complexity, which is crucial for robust solutions that perform well in real-world scenarios and generalize across diverse visual environments. Additionally, we evaluate the performance, considering both quality enhancement and runtime, of two handcrafted and two \ac{ml}-based \ac{sr} models on the validation and testing sets of \ac{vista}.
 \\ 
\textbf{Dataset URL:} \href{https://github.com/Omnidirectional-video-group/ODVista}{https://github.com/Omnidirectional-video-group/\\ODVista} 
\end{abstract}

\begin{keywords}
Omnidirectional video, 360-degree video,  super-resolution, video streaming, virtual reality,  machine learning. 
\end{keywords}

\acresetall

\section{Introduction}
Advancements in immersive video technologies have paved the way for users to engage in a virtual environment that mirrors real-life scenarios, thereby enhancing user engagement and a sense of belonging in a digital space. Various visual media formats, including \ac{odv}, volumetric videos, and light fields, are common and effective methods for facilitating an immersive viewing experience. In particular, \ac{odv}, also known as 360-degree video, has gained widespread popularity due to the availability of acquisition and display devices, as well as standardization efforts to ensure interoperability. However, the adoption of these videos in streaming encounters challenges related to bandwidth and latency, particularly in mobility conditions such as \acp{uav}. Adaptive resolution and compression aim to maintain quality while minimizing latency in such scenarios. Nevertheless, the process of downscaling and encoding may result in quality degradation and the introduction of compression artifacts. 

\begin{figure*}[t]
\centering
 \includegraphics[scale=0.55]{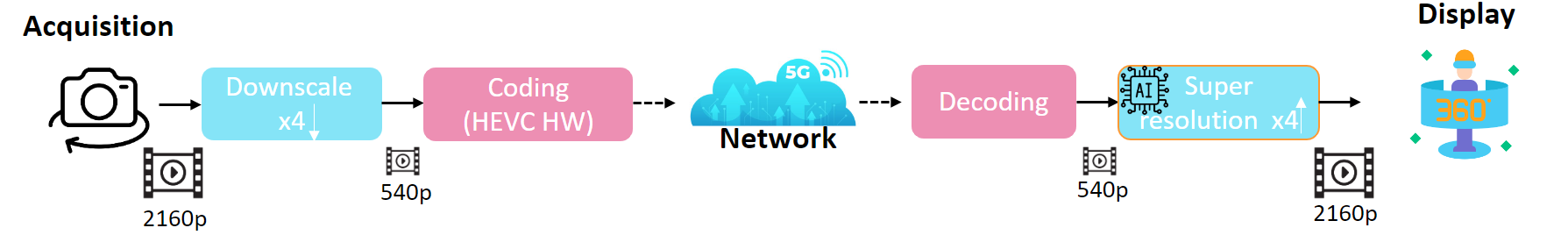}
\caption{Super-resolution integration in a typical streaming pipeline.}
\label{pipeline}
\end{figure*}

On the other hand, studies have demonstrated significant advancements in image and video \ac{sr} tasks through the adoption of deep learning-based methods. These methods, particularly those utilizing \ac{cnn}\cite{10.1007/978-3-319-10593-2_13}, \ac{vit}\cite{9607618, 10204527}, \ac{gan}\cite{8099502}, and \ac{rnn}\cite{9730760}, have pushed the boundaries of what can be achieved in terms of image clarity and detail enhancement. The deep learning models are trained on vast datasets of low-resolution and high-resolution image pairs, enabling them to learn complex mappings between the two. Additionally, in the typical \ac{odv} video streaming pipeline, as illustrated in Fig.~\ref{pipeline}, \ac{sr} can be integrated to enable effective upsampling of videos from lower resolutions to higher resolutions. In some cases, the original video can be intentionally downscaled to a lower resolution to preserve bandwidth. At the receiving end, the \ac{sr} algorithm is applied to upscale the video back to a superior resolution. This process significantly enhances the viewing experience by providing higher resolution video without the need for increased bandwidth for direct high-resolution video streaming.

In the literature, there are datasets and training methodologies that enable \ac{sr} models to produce high-resolution content with notable accuracy. However, the shortage of high-quality \ac{odv} video datasets limits the advancement of the accuracy of different \ac{sr} models. Several proposals have introduced diverse datasets featuring distinct properties. Table~\ref{datasets} summarizes existing datasets with different characteristics. In the 2023 NTIRE challenge on $360^{\circ}$ \ac{sr}, Cao \etal \cite{cao2023ntire} introduced a significant video dataset named \textit{ODV360}. This dataset contains high-resolution (2K) $360^{\circ}$ video content, featuring a total of 210 videos. The collection includes 90 videos from YouTube \cite{youtube} and existing public $360^{\circ}$ video datasets, alongside 120 videos directly recorded with Insta $360^{\circ}$ cameras. In~\cite{8019351}, Xu \etal proposed a dataset of 48 \ac{odv} sequences, each showcasing a wide variety of content that allows for categorization based on the video content. These sequences have been sourced from YouTube and other public domains under a free-use license. Subsequently, the original videos were edited to create short clips, with lengths ranging from 20 to 60 seconds. The video resolutions range from 3K (2880$\times$1440) up to 8K (7680$\times$3840), ensuring a broad range of details. Although this dataset is proposed for subjective quality assessment purposes, it can be very useful for \ac{sr} tasks due to its variety and high-quality content. Li \etal in~\cite{Li:2018:BGV:3240508.3240581} proposed a dataset that features 600 \ac{odv} sequences, including 60 high-quality reference sequences with diverse content, sourced from raw formats and YouTube's \ac{vr} channel at bitrates exceeding 15 Mbps. These reference sequences span various content categories like nature, shows, and sports, with resolutions ranging from 4K to 8K. Additionally, these sequences are edited to lengths of 10 to 23 seconds at frame rates of 24-30 \ac{fps} and are organized into 10 groups to enhance diversity and facilitate subjective analysis, ensuring varied resolution and content across groups. While these datasets exhibit commendable variety across multiple dimensions, including content diversity, resolution, frame rate, and, in certain cases, a broad spectrum of bitrates, they fall short in a crucial area: compression distortion. This aspect plays a crucial role in developing \ac{sr} models for streaming applications, as it introduces a loss of quality resulting from the video compression process. This distortion adds an extra layer of degradation, secondary to downscaling, affecting the visual fidelity of content in practical scenarios such as live streaming and \ac{vod} services. Addressing this challenge is essential for enhancing \ac{sr} model performance in real-world streaming environments. The absence of the compression factor in existing datasets limits the ability to fully improve upon how \ac{sr} algorithms can adapt to the artifacts introduced by compression. In this study, we propose \ac{vista}, a comprehensive \ac{odv} dataset designed specifically to address \ac{sr} challenges in the context of video streaming. The main contributions of this paper can be summarized as follows:

\begin{itemize}
    \item Introducing \ac{vista}, a novel dataset (Tables~\ref{datasets}) featuring a diverse array of scenes with both scaling (2$\times$ and 4$\times$) and compression distortions, aimed at facilitating the development of advanced \ac{sr} models for streaming scenarios.
    \item Adopting a balanced sampling strategy based on spatial and temporal complexity, to ensure a balanced distribution and reduces outliers, enhancing model robustness.
    \item Propose a novel evaluation metric that can effectively capture the trade-off performance of \ac{sr} techniques, specifically in balancing the enhancement of quality and runtime processing.   
    \item Evaluating the efficacy of various \ac{sr} approaches, including conventional and \ac{ml}-driven methods, establishing a benchmark for future research.
\end{itemize}

The rest of the paper is organized as follows. Section \ref{proposed-dataset} presents the proposed dataset with video collection and characteristics details. Section~\ref{benchmark} evaluates \ac{sr} algorithms, both traditional and \ac{ml}-based, on the \textit{ODVista} dataset to showcase its effectiveness across different baseline models. Finally, Section~\ref{conc} concludes the paper.  

\begin{table*}[t]
\centering
\caption{Summary of existing 360-degree video datasets.}
\adjustbox{max width=\textwidth}{
\begin{tabular}{lcccccccc}
\toprule
Database & Year &  \#Count\tnote{a} &  \#Total\tnote{b} & Resolutions   & \#Frames & Distortion type &  Standard (Encoder) \\
\midrule
%Flickr360 \cite{cao2023ntire} & 2023  & 3150 &  &  &  &  &  &   \\ image dataset
ODV360~\cite{cao2023ntire} & 2023 &  210 & 630 &  2K & 100  & scaling ($\times$2, $\times$4) & \xmark \\
VQA dataset~\cite{8019351} & 2017 & 48 & 48 & 3K, 8K  & 600-1800 &  \xmark  & \xmark \\
VQA-ODV~\cite{Li:2018:BGV:3240508.3240581} &  2018 & 600 & 600 & 4K & 240-690  & \xmark & \xmark \\

%360 Video DASH~\cite{10.1145/3587819.3592548} & 2023 &  9 & 9 & 4K & 2160-8790 & - & \xmark \\
%CVIQD2018 \cite{8547102} & 2018 &  &  &  &  &  & \\ image dataset
ODVista (ours) & 2024 & 200 & 1600 & 2K, 4K & 100  & scaling ($\times$2, $\times$4) \& compression & \acs{hevc}~\cite{6316136} (hevc\_nvenc) \\
\bottomrule
\end{tabular}
}
\begin{tablenotes}
\footnotesize
\item \#Count: Total number of unique contents. \#Total: Total number of video sequences, including reference and distorted videos.
\end{tablenotes}

\label{datasets}
\vspace{-1.5mm}
\end{table*}

\begin{figure}[t]
\centering
 \includegraphics[width=0.99\linewidth]{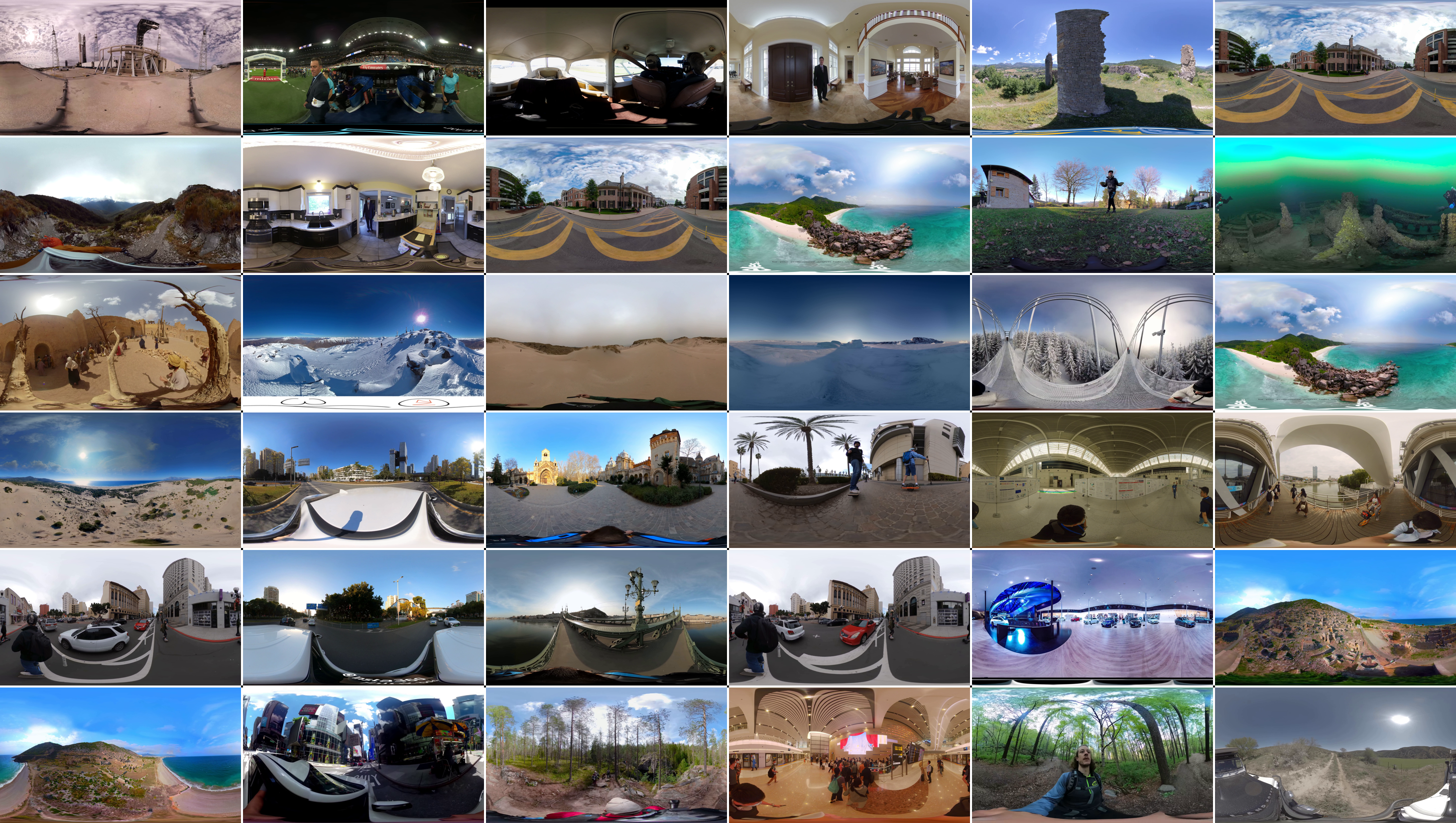}
\caption{Sample frames of the proposed \acs{vista} dataset.}
\label{frames}
\end{figure}

\section{ODVista dataset}
\label{proposed-dataset}
\subsection{Video collection}
In the realm of \ac{vr} and \ac{xr} research, a notable gap has been identified: the limited availability of diverse high-quality \ac{odv} datasets that are both comprehensive and open to the academic community. To address this limitation, we meticulously collected the \ac{vista} dataset, which includes 200 high-quality omnidirectional videos. These videos are equally divided into 100 videos with 2K (1080p) resolution and 100 videos with 4K (2160p) resolution, carefully collected from YouTube~\cite{youtube} and the ODV360 dataset~\cite{cao2023ntire}, ensuring the inclusion of only high-quality sequences. All videos are licensed under \ac{cc} for academic and research purposes. To ensure homogeneity, all sequences were subjected to a scene segmentation process, guaranteeing that each sequence comprised only a single scene. This segmentation was accomplished using the PySceneDetect tool~\cite{pydetect}. As a final step, all sequences were temporally cropped to 100 frames, if not originally formatted as such, resulting in a dataset that offers a comprehensive view of different scenes and scenarios. All these video sequences are stored in \ac{erp} format. The proposed dataset includes a variety of indoor and outdoor scenes, as well as dynamic sports content. This variety is crucial to simulate realistic scenarios. Sample frames from the \ac{vista} dataset are shown in Fig.~\ref{frames}, highlighting this diversity.

%In the realm of \ac{vr} and \ac{xr} research, a notable gap has been identified: the limited availability of diverse high-quality \ac{odv} datasets that are both comprehensive and open to the academic community. To address this limitation, we meticulously collected the \ac{vista} dataset, which includes 200 high-quality omnidirectional videos. These videos, equally divided into 100 videos with 2K resolution and 100 videos with 4K resolution, have been carefully collected from YouTube~\cite{youtube} and the ODV360 dataset~\cite{cao2023ntire}, ensuring the inclusion of only high-quality sequences. All videos are licensed under \ac{cc} for academic and research purposes. To ensure homogeneity, all sequences were subjected to a scene segmentation process, guaranteeing that each sequence comprised only a single scene. This segmentation was accomplished using the PySceneDetect tool~\cite{pydetect}. As a final step, all sequences were temporally cropped to 100 frames, if not originally
%formatted as such, resulting in a dataset that offers a comprehensive view of different scenes and scenarios. All these video sequences are stored in \ac{erp} format. The proposed dataset includes a variety of indoor and outdoor scenes, as well as dynamic sports content. This variety is crucial to simulate realistic scenarios. Sample frames from the \ac{vista} dataset are shown in Fig.~\ref{frames}, illustrating this diversity.  

\begin{figure*}[t]
\centering
\scriptsize
\centering
\begin{minipage}[b]{0.33\linewidth}
\centering
\centerline{\includegraphics[width=1\linewidth]{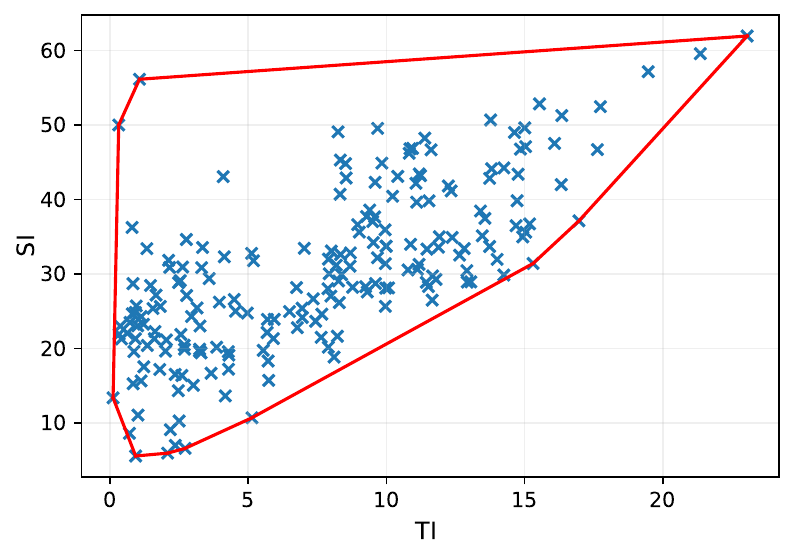}}
(a) SI vs. TI
\end{minipage}
\begin{minipage}[b]{0.33\linewidth}
\centering
\centerline{\includegraphics[width=1\linewidth]{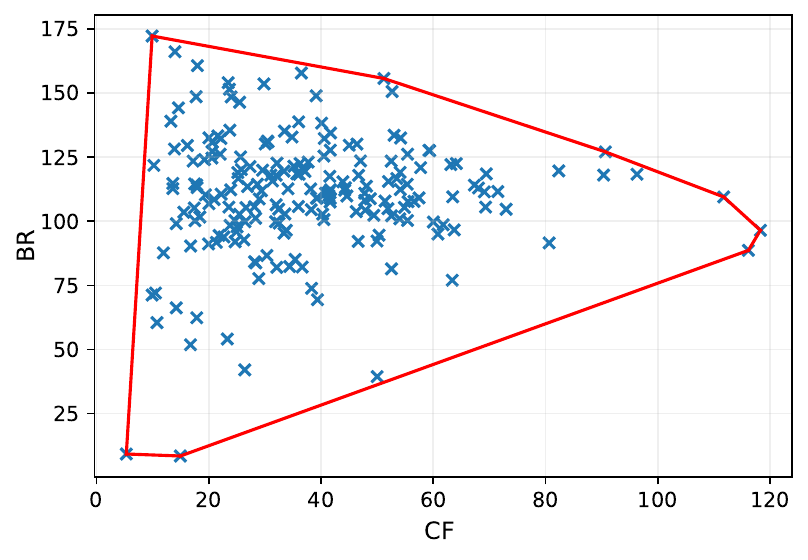}}
(b) BR vs. CF
\end{minipage}
\begin{minipage}[b]{0.33\linewidth}
\centering
\centerline{\includegraphics[width=1\linewidth]{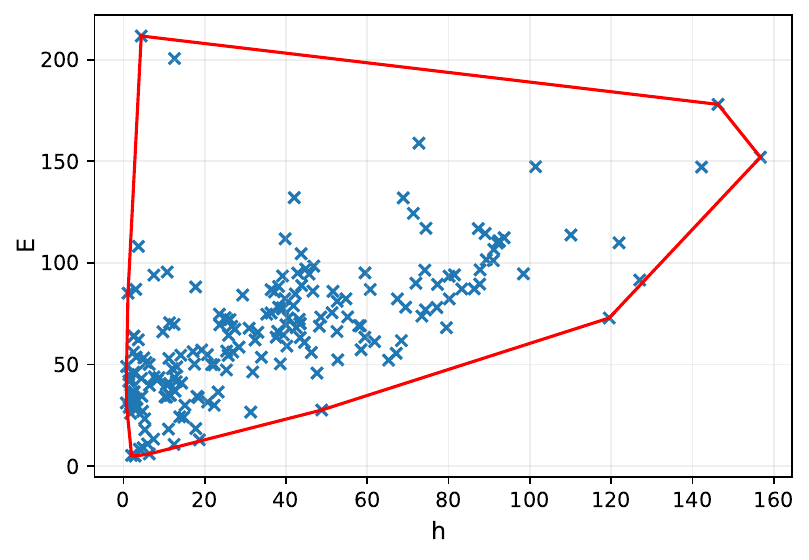}}
(b) h vs. E
\end{minipage}
\caption{Source content distribution in paired feature space with corresponding convex hulls. Left column: TI versus SI, middle column: CF versus BR and right column: h versus E.}
\label{features_plots}
\end{figure*}

\begin{figure}[t]
\centering
 \includegraphics[width=0.93\linewidth]{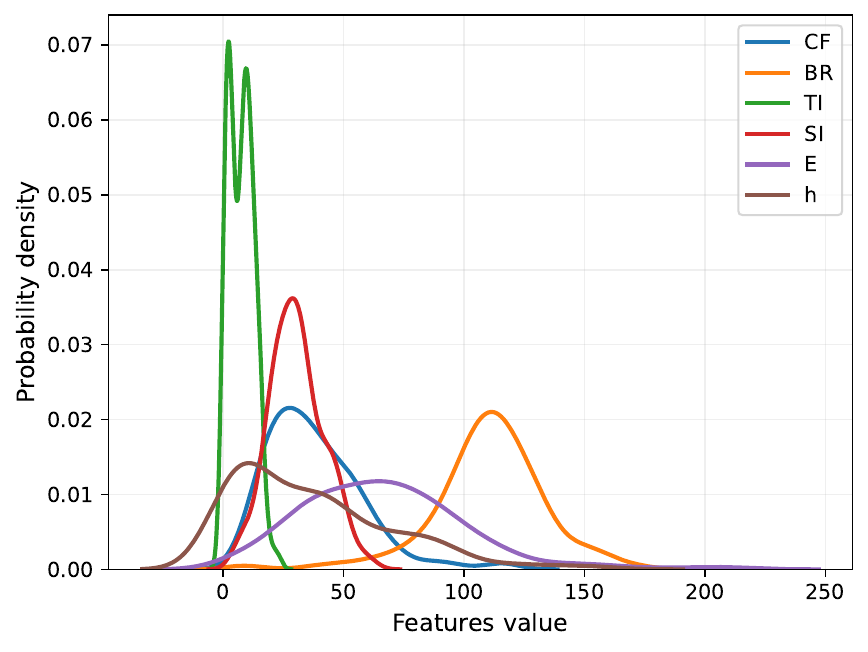}
\caption{Feature distribution comparisons among the proposed dataset.}
\label{distrib_features} 
\vspace{-1.5mm}
\end{figure}

\subsection{Dataset characterization}
As a means of characterizing the content diversity of the videos in databases, Winkler \etal \cite{6280595} initially proposed three video descriptors: spatial activity, temporal activity, and colorfulness. In our work, to achieve a more comprehensive analysis of content diversity, we expanded the set of these descriptors to include six low-level features, namely \ac{si}, \ac{ti}, \ac{br}, \ac{cf}, and two features derived from \ac{vca}~\cite{menon2022vca}, which are spatial complexity (E) and temporal complexity (h). Each of these features is calculated separately for each frame in the dataset. Subsequently, the computed values are averaged to obtain an overall (mean) representation. Scatter plots with convex hulls of paired features, illustrating the feature coverage of the proposed database, are shown in Fig.~\ref{features_plots}. Moreover, the fitted kernel distribution of each selected feature is illustrated in Fig.~\ref{distrib_features}. Firstly, we observe that video sequences cover a wide range in the spatiotemporal domain, with values ranging from 5 to 62 for \ac{si} and 0 to 24 for \ac{ti}, highlighting the diversity of the dataset. Furthermore, it is evident that the proposed dataset exhibits an extensive range of spatiotemporal complexities. The majority of the videos have low complexity as they contain only one scene, while a smaller portion consists of very complex sequences. Additionally, the scatter plot comparing \ac{br} to \ac{cf} reveals a diverse range of content types in the dataset, with sequences ranging from 10 to 174 in \ac{br} and from 0 to 120 in \ac{cf}. The range of \ac{br} values suggests the presence of various lighting conditions and scenes, spanning a variety of indoor and outdoor scenes, while the range of \ac{cf} values indicates a variety of color palettes and visual styles.

\subsection{Dataset processing}
\begin{figure}[t]
\centering
 \includegraphics[width=0.99\linewidth]{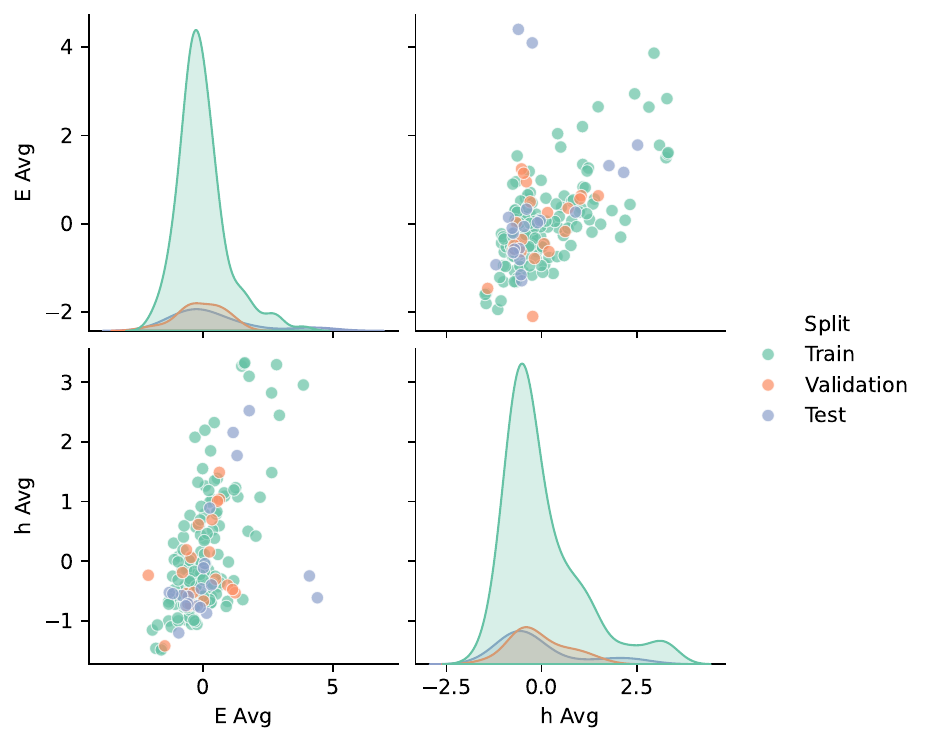}
\caption{Distributions of spatial complexity and temporal complexity across train, validation and test splits.}
\label{splits} 
\end{figure}

To construct a robust \ac{sr} dataset for \ac{odv} streaming scenarios, we employ two essential processing techniques: downscaling and compression. \\\\
{\bf Downscaling.} To generate the necessary \ac{lr} and \ac{hr} video pairs for \ac{sr} tasks, all sequences undergo downsizing using a Lanczos-3 filter~\cite{duchon1979lanczos}, as implemented by FFmpeg~\cite{ffmpeg}, at two different scales specifically, $\alpha=2$ and $\alpha=4$. This enables two distinct tracks for \ac{sr}: 2$\times$ and 4$\times$. \\\\
{\bf Compression.} Following the downscaling process, \ac{lr} sequences undergo compression. We utilize a hardware-based implementation of the \ac{hevc}/H.265 standard, embeded on an NVIDIA RTX A2000 8GB graphics card. The encoding process is carried out in \ac{ra} at four distinct low bitrates (0.25 Mbps, 0.5 Mbps, 1 Mbps, and 2 Mbps) to accurately simulate bandwidth constraints encountered in mobility scenarios, such as \acp{uav}. NVIDIA's implementation (NVENC~\cite{nvidia}) offers various presets for live streaming environments. Our evaluations determined that the "low latency high quality" (llhq) preset provides the best compromise, ensuring high-quality output while meeting the real-time constraints critical to streaming applications in energy-aware devices and  dynamic bandwidth environments. Consequently, $8$ different bitstreams are encoded for each source content, resulting in a total of $1600$ encoded video sequences. \\\\
{\bf Data splitting.}
In contrast to the conventional approach of random splitting employed by the majority of datasets, we divide our data into distinct sets, 80\% for training, 10\% for validation, and 10\% for testing, using the stratified sampling technique. This stratification relies on a K-means clustering approach, focusing on spatial and temporal complexity features (E and h), specifically the mean and average across frames. The optimal number of clusters (k) is determined using the Elbow method~\cite{kodinariya2013review}. This splitting strategy ensures a balanced distribution, effectively minimizing outliers, and enables the reliable and robust development of \ac{sr} methods. Fig.~\ref{splits} illustrates the distribution of spatial and temporal complexity across different splits, indicating well-balanced partitions.

\section{Benchmark} 
\label{benchmark}
In this section, we evaluate the performance of various \ac{sr} techniques, including both conventional and \ac{ml}-based methods, on the proposed \ac{vista} dataset.

\subsection{Baselines}
\subsubsection{Conventional methods}
\textbf{Bicubic interpolation~\cite{keys1981cubic}.} Bicubic interpolation is a commonly used method for image scaling in conventional \ac{sr} techniques. It calculates the values of new pixels by applying a weighted average to the 16 surrounding pixels within a 4$\times$4 neighborhood. \\\\
\textbf{Lanczos filter~\cite{duchon1979lanczos}.} Lanczos filter is an advanced conventional \ac{sr} method. It uses a sinc-based kernel, called Lanczos kernel, for estimating pixel values. Unlike bicubic interpolation, which uses a 4$\times$4 pixel grid, the Lanczos filter can take into account a larger number of surrounding pixels, with the exact number depending on the kernel size (e.g., Lanczos-3, Lanczos-4).

In our benchmark, we used the implementation provided by OpenCV~\cite{opencv} for both conventional methods. Specifically, for the Lanczos filter, we employ a kernel size of 4.

\subsubsection{AI-based methods}
\textbf{FSRCNN \cite{dong2016accelerating}.} \Ac{fsrcnn} is a \ac{cnn}-based method, evolved from SRCNN~\cite{dong2015image}, designed for real-time \ac{sr} applications. It features a more compact architecture that directly processes \ac{lr} inputs, thereby reducing computational complexity. \ac{fsrcnn}'s notable performance is attributed to the use of deconvolution layers towards the end of the network, enabling it to enlarge the image size and reduce processing time in a single step. \\\\
\textbf{SwinIR \cite{liang2021swinir}.} SwinIR, short for Image Restoration using Swin Transformer, is a transformer-based model designed for tasks such as \ac{sr}, image denoising, and compression artifact reduction. Derived from the Swin Transformer architecture~\cite{liu2021swin}, SwinIR processes images at various scales through a unique shifted windowing scheme for self-attention. This design allows SwinIR to detect and interpret complex image patterns and dependencies over long distances more efficiently than traditional convolutional methods.

\begin{table*}[]
\centering
\caption{Performance comparison of evaluated super-resolution methods.}
\label{perf}
\begin{tabular}{@{}llccccccc@{}}
\toprule
\multirow{2}{*}{Scale} & \multirow{2}{*}{Baseline} & \multicolumn{2}{c}{Validation set} & \multicolumn{2}{c}{Test set} & \multirow{2}{*}{\begin{tabular}[c]{@{}c@{}}Runtime/\\ 2k frame (s) $\downarrow$\end{tabular}} & \multirow{2}{*}{\begin{tabular}[c]{@{}c@{}}Runtime/\\ 4k frame (s) $\downarrow$\end{tabular}} & \multirow{2}{*}{$Q$ $\uparrow$} \\ \cmidrule(lr){3-6}
 &  & \acs{ws-psnr} (dB) $\uparrow$ & \acs{ws-ssim} $\uparrow$ & \acs{ws-psnr} (dB) $\uparrow$ & \acs{ws-ssim} $\uparrow$ &  &  &  \\ \midrule
\multirow{4}{*}{$2\times$} & SwinIR{\color{red}$ ^\maltese$} & 29.664 & 0.8437 & 29.761 & 0.8250 & 1.5232  & 7.5360  & 27.29 \\
 & \acs{fsrcnn}{\color{red}$ ^\maltese$} & 29.113 & 0.8321 & 29.280 & 0.8149 & 0.0015 & 0.0009  & 66.51 \\
 & Bicubic{\color{green} $^\clubsuit$} & 28.829 & 0.8060 & 28.743 & 0.8117 & \xmark & \xmark & \xmark \\
 & Lanczos~4{\color{green} $^\clubsuit$} & 28.880 & 0.8064 & 28.797 & 0.8110 & \xmark & \xmark & \xmark \\ \midrule
\multirow{4}{*}{$4\times$} & SwinIR{\color{red}$ ^\maltese$} & 28.811 & 0.8313 & 29.065 & 0.8099 & 0.4458  & 1.5155  & 29.79 \\
 & \acs{fsrcnn}{\color{red}$ ^\maltese$} & 28.018 & 0.8107 & 28.317 & 0.7912 & 0.0013  & 0.0015  & 61.10 \\
 & Bicubic{\color{green} $^\clubsuit$} & 27.585 & 0.7982 & 27.790 & 0.7831 & \xmark & \xmark & \xmark \\
 & Lanczos~4{\color{green} $^\clubsuit$} & 27.860 & 0.7995 & 27.795 & 0.7814 & \xmark & \xmark & \xmark \\ \bottomrule
\end{tabular}%
{\begin{flushleft}
 {\color{red}$ ^\maltese$} \acs{ml}-based \acs{sr} methods, {\color{green} $^\clubsuit$} Handcrafted-based \acs{sr} methods.
   \end{flushleft}}
\end{table*}

\subsection{Evaluation metrics}
To assess the performance of the baseline models on the proposed dataset, we evaluate both the quality of the upscaled video and the complexity of the \ac{sr} process. Quality assessment is carried out using the objective quality metrics specifically designed for 360-degree videos, namely \ac{ws-psnr} and \ac{ws-ssim}. The complexity of the model performing \ac{sr} is measured by the inference time on a PC fitted with an Intel® Xeon 8280 CPU @ 2.70GHz × 56, 128GB RAM, and a 48GB VRAM NVIDIA RTX 6000 Ada graphics card. In particular, the runtime of \ac{ml}-based models is estimated on the \ac{gpu} device. In order to evaluate the trade-off between quality enhancement and runtime, we propose a new scoring metric that takes into account both quality enhancement and runtime, as follows:
\begin{figure}[t]
\centering
 \includegraphics[width=0.99\linewidth]{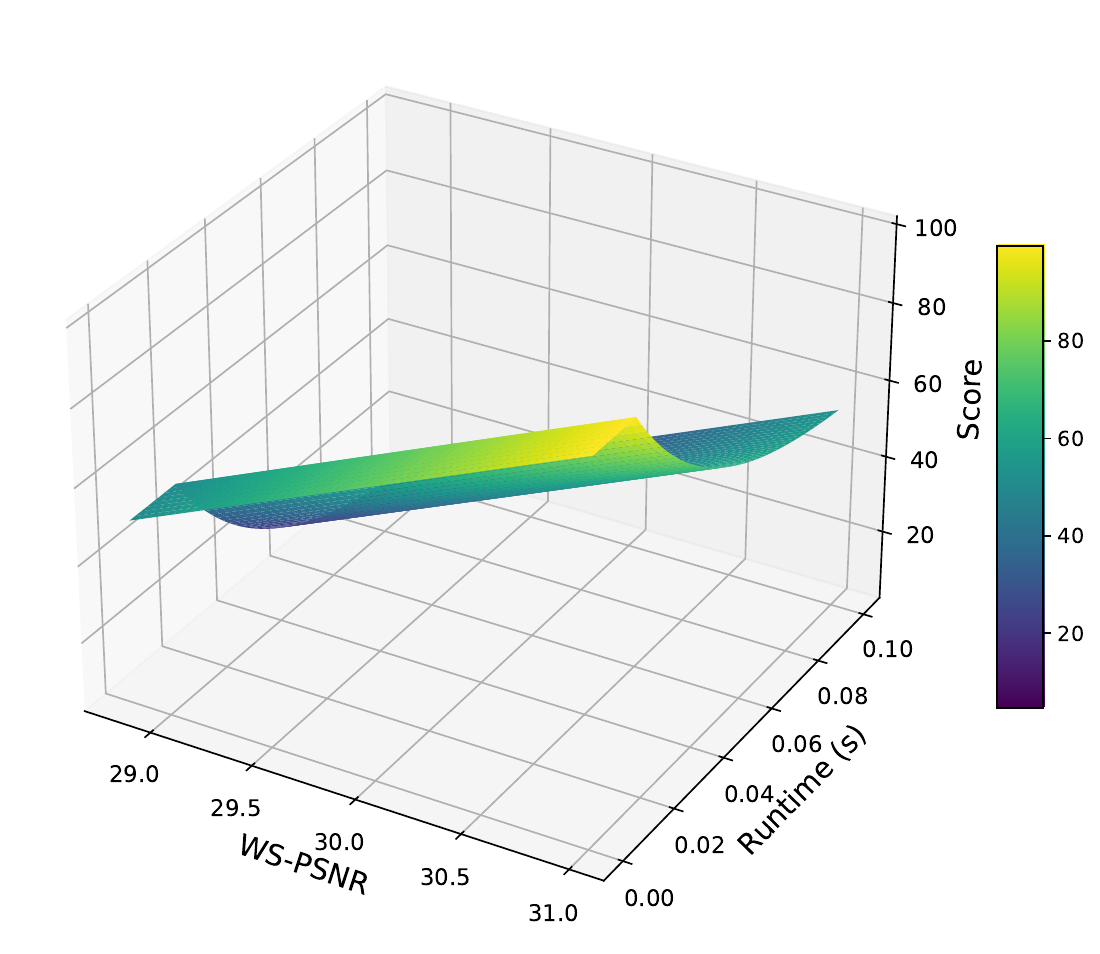}
\caption{3D Visualization of score metric $Q$ variation with \acs{ws-psnr} (dB) and runtime (s), with $\beta$ = 0.5, $\acs{ws-psnr}_{min}$ = 28.8 dB and $\acs{ws-psnr}_{max}$ = 31 dB.}
\label{score}
\end{figure}

\begin{equation}
Q = (\beta \times \hat{Q} + (1 - \beta) \times C) \times 100,
\end{equation}
where $\beta$ is a weighting parameter (set to 0.5 in our evaluation), $\hat{Q}$ is the normalized \ac{ws-psnr} score, and $C$ is the runtime evaluation score. The normalized quality score $\hat{Q}$ is computed as:
\begin{equation}
 \hat{Q} =   \frac{\ac{ws-psnr}-\ac{ws-psnr}_{min}}{\ac{ws-psnr}_{max} -\ac{ws-psnr}_{min}},
\end{equation}
where $\text{WS-PSNR}_{\min}$ represents minimum value reached by the least performing model (i.e., Bicubic) and $\text{WS-PSNR}_{\max}$ defines the theoretical maximum \ac{ws-psnr} values we consider in our evaluation (30 dB and 31 dB for $\alpha=4$ and $\alpha=2$, respectively). The runtime evaluation metric assigns a full score to models that achieve a processing time of 0.016 seconds or less per 2K frame, as this speed enables a smooth 60 \ac{fps}, essential for high-quality live streaming. To reinforce this standard, we apply penalties in our evaluation criteria for runtimes that exceed 0.016 seconds. The slower the model, the larger the penalties as follows: 
\begin{equation}
C =
  \begin{cases}
    1  & \quad \text{runtime } \leq 0.016, \\
   e^{B \times (0.016 -\text{runtime})}  & \quad \text{otherwise,} \text{ with } B=30.
  \end{cases}
\end{equation}

%%%%%%%%%%%%%%%%%%%%%%%%%%%%%%%%%%%%%%%%%%%%%%%%%%%%%%%%%%%%%%%%
\iffalse

\begin{equation}
 Q =   ( \alpha \times  \hat{Q} + (1-\alpha) \times C) \times 100, 
\end{equation}
where $\alpha$ is a weighting parameter. $\hat{Q}$ is a normalized \ac{ws-psnr} score of the model compared to the minimum \ac{ws-psnr} value, reached by the least performing model (i.e., Bicubic).
\begin{equation}
 \hat{Q} =   \frac{\ac{ws-psnr}-\ac{ws-psnr}_{min}}{\ac{ws-psnr}_{max} -\ac{ws-psnr}_{min}}
\end{equation}
The runtime evaluation metric assigns a full score to models that achieve a processing time of 0.016 seconds or less per 2K frame, as this speed enables a smooth 60 \ac{fps}, essential for high-quality live streaming. To reinforce this standard, we apply penalties in our evaluation criteria for infernece times that exceed 0.016 seconds. The slower the model, the larger the penalties as follows: 
\begin{equation}
C =
  \begin{cases}
    1  & \quad \text{runtime } \leq 0.016 \\
   e^{B \times (0.016 -\text{runtime})}  & \quad \text{otherwise,} \text{ with } B=30
  \end{cases}
\end{equation}
\fi
%%%%%%%%%%%%%%%%%%%%%%%%%%%%%%%%%%%%%%%%%%%%%%%%%%%%%%%%%%%%%%%%

Fig.~\ref{score} illustrates the variation of the score $Q$ based on runtime and \ac{ws-psnr}. It is evident that the highest scores are achieved by models operating in real time (60 \ac{fps}) while delivering maximum \ac{ws-psnr}. Then, the score is penalized when there is an increase in runtime beyond real-time or when the model does not significantly outperform the quality of the baseline model.

%To assess the performance of the baseline models we consider both quality of the of the output video and the complexity of the \ac{sr} process. The quality is measured with the objective quality metric specifically designed for 360-degree video, namely \ac{ws-psnr}. The \ac{sr} complexity \ac{ml}-based models is estimated by the run time on \ac{gpu} NVIDIA A6000 device. To assess the trade-off between the quality and the run time, we propose a new score metric that consider both quality enhancement and run time as follows.     
\subsection{Results and analysis}
In this section, we assess the performance of the four baseline models on the validation and test sets of our proposed dataset, \ac{vista}. Table~\ref{perf} presents the results of the four baseline models introduced in Section~\ref{benchmark}, considering \ac{ws-psnr}, \ac{ws-ssim}, runtime, and $Q$ score. It is evident that \ac{ml}-based models significantly enhance the quality of the output videos. Specifically, there is an improvement of 0.78 dB and 0.23 dB in terms of \ac{ws-psnr} for the SwinIR and \ac{fsrcnn} models, respectively, compared to the best-performing handcrafted model, Lanczos~4, in the 2$\times$ scaling configuration. This improvement is even more pronounced, particularly for the SwinIR model, achieving a 0.95 dB higher \ac{ws-psnr} compared to Lanczos~4 in the 4$\times$ scaling configuration. These results are corroborated by the \ac{ws-ssim} metric.

However, the quality improvements brought by the SwinIR model come at the expense of higher complexity, requiring an average of 0.44 seconds to process one 2K resolution frame and 1.51 seconds for a 4K frame in the 4$\times$ scale, falling short of maintaining real-time processing. This latency is even higher (1.52 seconds to process one 2K resolution and 7.53 seconds for a 4K resolution) in the 2$\times$ scaling configuration, mainly caused by the higher resolution of the input video compared to the 4$\times$ scale. On the contrary, the \ac{fsrcnn} model exhibits a noteworthy balance between enhancing video quality and runtime efficiency. This model demonstrates the capacity to uphold real-time processing, surpassing 30 \ac{fps} in both scaling configurations, even at a 4K video resolution. The proposed metric, denoted as $Q$, underscores the superiority of the \ac{fsrcnn} model, securing the top rank (best trade-off between quality enhancement and runtime), and is subsequently trailed by the SwinIR model.

\section{Conclusion} 
\label{conc}
In this paper, we introduce the \ac{vista} dataset, designed to address compression and scaling distortions in \ac{odv}. The proposed dataset comprises 200 high-quality and high-resolution videos. Each video has been scaled by two distinct factors and encoded across four different low-bitrate ranges to accurately simulate real-world scenarios characterized by limited bandwidth conditions. The \ac{vista} dataset is characterized by its diversity, incorporating a range of indoor and outdoor scenes, covering various visual contents and distortion levels, making it well-suited for developing robust \ac{sr} models. Additionally, we utilized a stratified sampling technique to ensure balanced training, validation, and test sets, improving the representativeness of the dataset and facilitating efficient model training and evaluation. Furthermore, we provide a comprehensive benchmark to evaluate both conventional and \ac{ml}-based SR methods on the proposed dataset, enhancing its utility for the research community. As future work, we plan to expand the proposed dataset by including other videos captured using the Insta360 Pro 2 camera, further enriching the dataset. Furthermore, we aim to compress the dataset using additional video codecs that represent different video coding standards and formats, such as \ac{vvc}/H.266 and AV1.

%In this paper, we introduced the \ac{vista} dataset, designed to address compression and scaling distortions in \ac{odv}. The proposed dataset comprises 200 high quality and high resolution videos. Each video has been scaled by two distinct factors and encoded across four different low-bitrate ranges, to accurately simulate real-world scenarios characterized by limited bandwidth conditions. \ac{vista} dataset is characterized by its diversity, incorporating a range of indoor and outdoor scenes, covering various visual contents and distortion levels, making it well-suited for developing robust \ac{sr} models. Additionally, we utilized a stratified sampling technique,To ensure balanced train, validation and test sets, which improves the representativeness of the dataset and facilitates efficient model training and evaluation. Furthermore, we provide a comprehensive benchmark to evaluate some conventional and \ac{ml}-based SR methods on the proposed dataset, enhancing its utility for the research community. As future work, we plan to expand the proposed dataset by including other videos captured using the Insta360 Pro 2 camera, further enriching the dataset.

%\vfill\pagebreak
%\newpage

% References should be produced using the bibtex program from suitable
% BiBTeX files (here: strings, refs, manuals). The IEEEbib.bst bibliography
% style file from IEEE produces unsorted bibliography list.
% -------------------------------------------------------------------------

\bibliographystyle{IEEE}
\bibliography{bibfile}

\end{document}